\documentclass[twoside,a4paper]{article}
\usepackage{amssymb}
\usepackage{fleqn}
\usepackage{epsfig,a4}

\let\kappa=\varkappa

\def\text#1{\mbox{#1}}

\def\restylefloat#1{}
\newcommand{\ds}{\displaystyle}
\newcommand{\abc}[1]{\mbox{#1)}\quad}

\newcommand{\bm}[1]{\mbox{\boldmath $#1$}}

\newcommand{\deriv}[2]{\mbox{$\displaystyle
\frac{\mathrm{d}\,\! #1}{\mathrm{d}\,\! #2}$}}

\begin{document}  \thispagestyle{plain}

\title{Heat expansion of star-like cosmic objects induced by the cosmological
expansion\thanks{Dedicated to Prof. Dr. R. Kippenhahn on the
occasion of his 75th birthday}}
\author{E. Schmutzer, Jena, Germany \\
%EndAName
Friedrich Schiller University}
\date{Received: 2001}
\maketitle
\date{}

\begin{abstract}
The heat expansion of a star-like cosmic object, induced by the cosmological
bremsheat production within a moving body, that was predicted by the
Projective Unified Field Theory of the author, is approximately treated. The
difference to planet-like bodies investigated previously arises from another
material constitution.The exponential-like expansion law is applied to a
model with numerical values of the Sun. The results are not in contradiction
to empirical facts.
\end{abstract}

\section{Review}

These days it is generally recognized in science that, as the Sun burns
through its hydrogen on the main sequence, it steadily grows hotter and
therefore more luminous. It is also accepted that through this fact the Sun
continuoously grows larger. Recently an interesting paper on this subject
has been published (Korycansky et al. 2001).

This paper devoted to a similar subject follows the theoretical line of our
previous publication on the bremsheat expansion of a moving spherical
planet-like body, induced by the expansion of the cosmos (Schmutzer 2000a).
This hypothetical bremsheat effect has been derived from the Projective
Unified Field Theory (PUFT) of the author (Schmutzer 1995). The moving
star-like body considered here may be orbiting around a galactic central
body. As usual in astrophysics, the Gauss system of units is used. Let us
first repeat the general theoretical results.

The radial expansion velocity of the surface of the sphere ($r$ radius of
the sphere) reads
\begin{equation}
v=\deriv{r}{t}=4E_{S}r\sigma ^{3}\deriv{\sigma }{t}\, .
\label{eins}
\end{equation}
The quantities occurring mean (in our specific terminology):
$\sigma $ is the scalaric world function determined by the
scalaric cosmological field equation, whereas
\begin{equation}
E_{S}=\frac{\Sigma _{S}}{4}\left( \frac{\gamma _{N}M_{c}}{\sigma
F_{0}} \right) ^{2}
 \label{zwei}
\end{equation}
is the scalaric cosmic expansion factor with following meaning of
the symbols appearing: $\gamma _{N}=6.68\cdot 10^{-8}\mathrm{
g^{-1}\, cm^{3}\, s}^{-2}$ is the Newtonian gravitational constant
and $M_{c}$ the mass of the central
body. In this context one should take account that the relation $M_{c}=%
\mathcal{M}_{c}\sigma $ holds, where $\mathcal{M}_{c}$ is the
scalmass (invariant constant mass) of the central body. Further
\begin{equation}
\Sigma _{S}=\frac{f_{S}\alpha _{c}}{3c_{Q}}  \label{drei}
\end{equation}
is the scalaric material factor ($\alpha _{c}$ cubic thermal
coefficient of dilation, $c_{Q}$ specific heat, $f_{S}$ heat
consumption factor being an individual quantity of the order of
magnitude 1), and $F_{0}=\sigma |\bm{R\times V}|$ is the modified
arial velocity of the orbiting body (constant of integration)
appearing in its angular momentum ($\bm{V}$ orbital velocity, $|
\bm{R}| $ distance between the moving body and the galactic
central body).

For further mathematical treatment of the equation (\ref{eins}) we assume
constancy of the scalaric cosmic expansion factor (\ref{zwei}). Then we are
able to integrate with the result

\begin{equation}
r=r_{b}\exp [E_{S}(\sigma ^{4}-\sigma _{b}^{4})]\, ,
 \label{vier}
\end{equation}
where
\begin{equation}
\abc{a}
r_{b}=r(t=t_{b})\quad\text{and}\quad\abc{b}\sigma_{b}=\sigma
(t=t_{b}) \label{fuenf}
\end{equation}
mean the value of the radius of the sphere and the value of the scalaric
world function at the time of the birth of the cosmic object considered
(index $b$ refers to birth).

\section{Application of the theory to a star-like cosmic object (Sun as
model)}

For rough application of the theory presented above we refer to approximate
numerical values of the Sun as a numerical model (further short: sun)
orbiting around the center of the Galaxy (index $p$ refers to present, year
= y):

\begin{equation} \label{sechs}
\begin{array}{llll}\bigskip
\abc{a}R &=&2.6\cdot 10^{22}\, \mathrm{cm} &\text{(radius of the
orbit),}
 \\ \bigskip
 \abc{b} V &=&2.2\cdot 10^{7}\,\mathrm{cm \,s}^{-1}
 &\text{(orbital velocity of the sun),}
 \\ \bigskip
 \abc{c}
  M_{c} &=&1.8\cdot 10^{44}\,\mathrm{g} &\text{(acting mass of the Galaxy),}
\\ \bigskip
\abc{d} r_{p} &=&6.96\cdot 10^{10}\, \mathrm{cm}&\text{(present
radius of the sun),}
\\ \bigskip
\abc{e} t_{sun} &=&4.66\cdot 10^{9}\,\mathrm{y} &\text{(age of the
sun).}
\end{array}
\end{equation}
 Let us mention that the value of the central mass acting at
the place of the sun resulted from our theory of the rotation
curve (Schmutzer 2001). Obviously this value is different from the
usually accepted value of the mass of the whole Galaxy which is
somewhat greater.

Apart from these individual values (\ref{sechs}) we further need
the present cosmological values
\begin{equation}\label{sieben}
\begin{array}{lll}\bigskip
\abc{a}\sigma _{p} &=&65.19\, ,
  \\ \bigskip

\abc{b} \left( \deriv{\sigma}{t}\right)_{p} &=&1.14\cdot 10^{-9}\,
\mathrm{ y^{-1}\,.}
\end{array}
\end{equation}
 These values result from the solution of the system of the
cosmological differential equations of the closed isotropic
homogeneous cosmological model investigated in detail (Schmutzer
2000b).

By means of (\ref{sechs}a,b) and (\ref{sieben}a) we find the value
$ F_{0}=3.73\cdot 10^{31}\, \mathrm{cm^2\,s^{-1}}$.

The main problem arising now is to find a way to get information on the
scalaric expansion factor $E_{S}$ of the sun (\ref{zwei}). For comparing it
is appropriate to remember the results obtained for planet-like cosmic
bodies (Schmutzer 2000). We used the following values applied in geophysics
for the earth:

\begin{equation}\label{acht}
\begin{array}{lll}
\abc{a}\alpha _{c} &=&6.84\cdot 10^{-5}\mathrm{K}^{-1}\, ,\bigskip
 \\
\abc{b}c_{Q} &=&1.5\cdot 10^{7}\,\mathrm{cm^2\,s^{-2}\,K^{-1}}\,.
\end{array}
\end{equation}
For $f_{S}=0.5$ we obtaind the values
\begin{equation}\label{neun}
\begin{array}{lll}
\abc{a}\Sigma _{S} &=&7.6\cdot 10^{-13}\,\mathrm{cm^{-2}\,s^{2}}\,
, \bigskip
\\
\abc{b} E_{S} &=&9.35\cdot 10^{-8}\, .
\end{array}
\end{equation}

The situation with respect to the sun is insofar qualitatively different
from the situation just mentioned, since the interior of the sun consists of
gas (mainly H, partly atomic burning to He) with a very high temperature.

For simplicity we approximately treat this problem by referring to
an perfect gas for which the relation
\begin{equation}
\overline{\varepsilon}=\frac{m_{0}}{2}\,\overline{u^{2}}=\mathrm{\frac{3kT}{2}}
\label{einsnull}
\end{equation}
is valid ($m_{0}$ mass of a gas particle, $u$ velocity of the gas
particles subjected to averaging, k Boltzmann constant, $T$
kinetic temperature). As it is well known, for a perfect gas the
formulas
\begin{equation}\label{einseins}
\begin{array}{lll}
\abc{a} \alpha _{c} &=&\ds
\frac{1}{T}=\frac{3\mathrm{k}}{m_{0}\overline{u^{2}}}\, ,
\bigskip
 \\
 \abc{b}c_{Q} &=&\ds \frac{3\mathrm{k}}{2m_{0}}
\end{array}
\end{equation}
hold. Hence the relations (\ref{drei}) and (\ref{zwei}) take the form
\begin{equation}
\Sigma _{S}=\frac{2f_{S}m_{0}}{3\mathrm{k}T}  \label{einszwei}
\end{equation}
and

\begin{equation}
E_{S}=\frac{f_{S}m_{0}}{6\mathrm{k}T}\left(\frac{\gamma
_{N}M_{c}}{\sigma F_{0}}\right) ^{2}\, .  \label{einsdrei}
\end{equation}
Using the numerical values (\ref{sechs}) and (\ref{sieben}) for
the sun, we find
\begin{equation}
\left( \frac{\gamma _{N}M_{c}}{\sigma F_{0}}\right) ^{2}=2.45\cdot
10^{7}\, \mathrm{cm^{2}\, s^{-2}}\,.  \label{einsvier}
\end{equation}
As mentioned above, it is our aim to treat cosmic bodies globally.
Therefore it seems to be acceptable to take the value $T=5\cdot
10^{6}\mathrm{\,K}$ as an appropriate mean value of the kinetic
temperature of the sun. Let us further remember the value of the
Boltzmann constant $\mathrm{k = 1.38 \cdot 10^{-16}\, g\, cm^{2} s
^{-2}\, K^{-1}}$ and the value of the proton mass $m_{p}=1.68\cdot
10^{-24}\,\mathrm{g}$ (being used as the mass of the H-atom
constituent of the gas
considered). Hence we obtain from (\ref{einsdrei}) the numerical result ($%
f_{S}=1)$
\begin{equation}
E_{S}=9.92\cdot 10^{-9}\,. \label{einsfuenf}
\end{equation}

\section{Numerical evaluation and plotting for the Sun (as a model)}

\subsection{Numerical formulas for radius and expansion velocity}

Inserting the values (\ref{einsfuenf}), (\ref{sechs}d) and (\ref{sieben})
into formula (\ref{eins}), we find for the present expansion velocity of the
surface of the sun
\begin{equation}
v_{p}=0.87\,\mathrm{cm\, y^{-1}}\,.  \label{einssechs}
\end{equation}

Further we get information from formula (\ref{vier}) on the
temporal behavior of the radius of the sun. Inserting
(\ref{einssechs}) into this equation we arrive at
\begin{equation}
r_{p}=r_{b}\exp [9.92\cdot 10^{-9}(\sigma _{p}^{4}-\sigma
_{b}^{4})]\,. \label{einssieben}
\end{equation}
Now from (\ref{sechs}d) we know the quantity $r_{p}$ and from
(\ref{sieben}a) the quantity $\sigma _{p}.$ As mentioned above,
there is a good consensus among astrophysicists that the age of
the sun is about 4.66 billion years. Therefore we need the value
of $\sigma $ at the time of birth of the sun. In order to find
this value a small excursus to the results of our cosmological
model quoted above is necessary.

We arrived at an age of this cosmos of $18\cdot
10^{9}\,\mathrm{y}$, i.e. we need the cosmological values at the
time $t_{b}=13.34\cdot 10^{9}\,\mathrm{y}$. Checking our list of
numerical data for the whole cosmos after the big start (Urstart),
we find the values $\sigma _{b}=\sigma (t=t_{b})=55.44$ and
$\sigma _{b}^{4}=9.45\cdot 10^{6}$.

Now we are prepared to calculate the radius of the sun at the birth with the
result
\begin{equation}
r_{b}=6.39\cdot 10^{10}\mathrm{cm\, .}  \label{einsacht}
\end{equation}

\subsection{Plotting of the temporal course of the radius and the velocity}

The abscissa of the figures presented in the following refers to
the rescaled time $\eta =0.945\cdot 10^{-9}t\mathrm{\,y^{-1}}$.
This practically means that the unit $\Delta \eta =1$ roughly
corresponds to $10^{9}\,\mathrm{y}$. Referring to the figures of
the radius, the unit of the radius at the ordinate is given in cm;
referring to the figures of the velocity, the unit of the velocity
at the ordinate is $\mathrm{cm/y}$.

The temporal course of the radius of the spherical body is
presented in Fig. 1 for the whole time scale since the big start.
\begin{figure}[hbt]
\begin{center}
\footnotesize
\includegraphics[width=0.75\columnwidth]{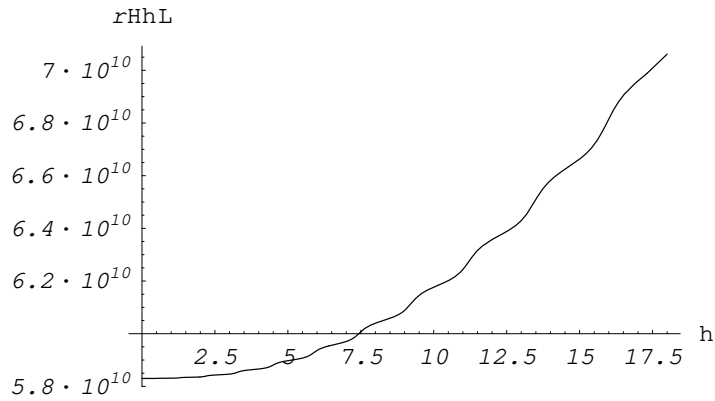}  % Filename.eps
\caption{Temporal course of the radius for the whole time scale
from the big start to $18\cdot 10^{9}\,\mathrm{y}$}
\label{fig:1}
\end{center}
\end{figure}

The temporal course of the radial velocity of the surface of the
spherical body is presented in Fig. 2 which shows the course of
the velocity for the whole time scale since the big start.
\begin{figure}[hbt]
\begin{center}
\footnotesize
\includegraphics[width=0.75\columnwidth]{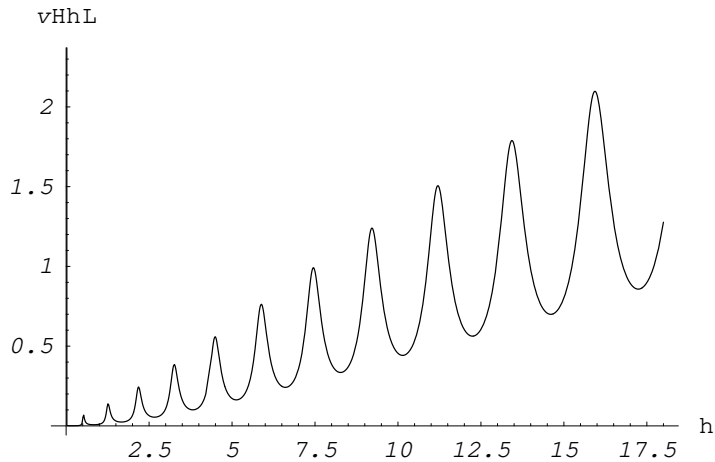}  % Filename.eps
\caption{Temporal course of the radial velocity of the surface for
the whole time scale from the big start to $18\cdot
10^{9}\,\mathrm{y}$} \label{fig:2}
\end{center}
\end{figure}

\subsection{Annotation to the mass of the Galaxy}

In investigating the orbital motion of the sun around the central
body of the Galaxy, for physical reasons it is necessary to use
the gravitational force just at the place of the moving sun.
Because of the matter (particularly dark matter) outside the sun
it is obvious that the mass causing the gravitational force
mentioned is not the total mass of the Galaxy. We recently treated
the problem of the radial mass distribution in the Galaxy within
the framework of our theory of the rotation curve of stars
(Schmutzer 2001). In that paper we came to the numerical value
(\ref{sechs}c) of the gravitationally acting mass at the place of
the sun. All numerical calculations performed above are based on
this value.

From formula (\ref{zwei}) we learn that the scalaric cosmic
expansion factor $E_{S}$ contains $M_{c}$ quadratically. That
means under the circumstances of the exponential character of the
expansion law (\ref{vier}) an extraordinary sensitivity of the
radius on the mass $M_{c}$.

With the aim of illustration we also performed the calculations on
the basis of the total mass of the Galaxy $M_{c/total}=2\cdot
10^{45}\,\mathrm{g}$ (to be found in astrophysical literature),
arriving at the rescaled temporal curves for the radius $R(\eta )$
and the radial expansion velocity $V(\eta )$, both presented in
Fig. 3 and Fig. 4. Hence the present value of the radial expansion
velocity $V_{p} =55\,\mathrm{cm/y}$ results.

The following figures show the temporal course of $R(\eta )$ and $V(\eta ).$
\begin{figure}[hbt]
\begin{center}
\footnotesize
\includegraphics[width=0.75\columnwidth]{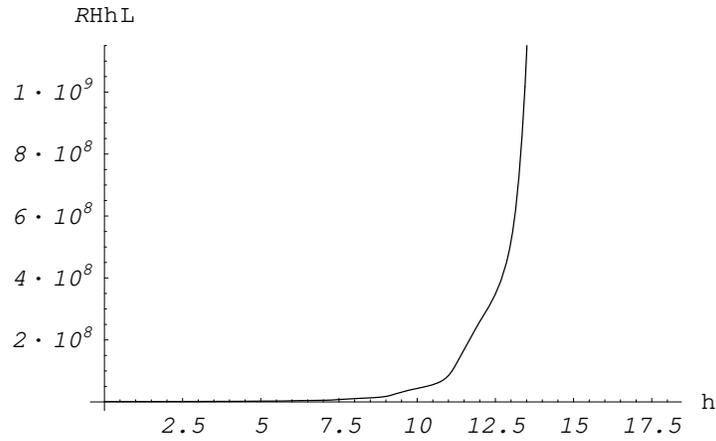}  % Filename.eps
\caption{Temporal course of the radius for the whole time scale
from the big start to $18\cdot 10^{9}\,\mathrm{y}$ (different
numerical variant)} \label{fig:3}
\end{center}
\end{figure}
\begin{figure}[hbt!]
\begin{center}
\footnotesize
\includegraphics[width=0.75\columnwidth]{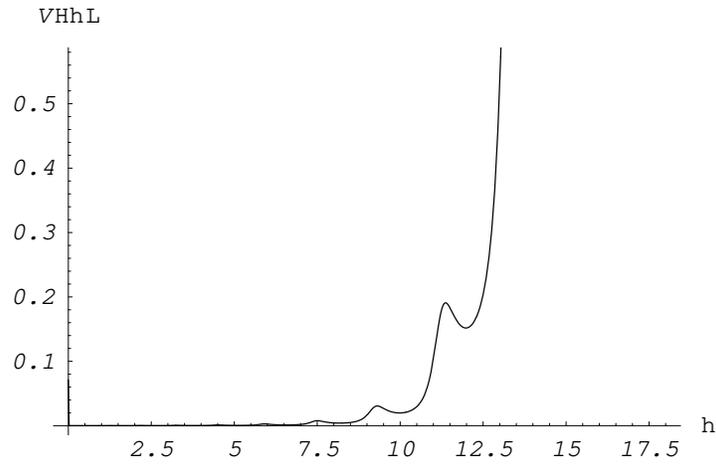}  % Filename.eps
\caption{Temporal course of the radial velocity of the surface for
the whole time scale from the big start to $18\cdot
10^{9}\,\mathrm{y}$ (different numerical variant)} \label{fig:4}
\end{center}
\end{figure}

Let us conclude this paper with the general remark that in Fig. 1
and Fig. 2 the curves show a quasi-continuous course through the
whole time scale from the big start to the presence, i.e. there is
no hint at the time of birth of the cosmic object considered. In
contrast to this behavior in Fig. 3 and Fig. 4 the birth is
significantly marked by a steep rise. All in all it seems that
this theory corresponds to a longtime evolution of the star-like
cosmic objects from ``small to large'' and not according to the
accretion concept from ``large to small''.

I would like to thank Prof. Dr. R.Kippenhahn (Goettingen) for
interesting in\-formation on the physics of the sun and Prof. Dr.
A.Gorbatsievich (Minsk) for advice and help. \vspace{2ex}

\noindent \textbf{References}
\begin{description}
\item[]
Korycansky, D. G. , G. Laughlin and Adams, F. C.: 2001,
arXiv:astro-ph/0102126 (7 Feb), Astrophysics and Space Science (in
press)
\item[]
Schmutzer, E.: 1995, Fortschritte der Physik 43, 613
\item[]
Schmutzer, E.: 2000a, Astron. Nachr. 321, 227
\item[]
Schmutzer, E.: 2000b, Astron. Nachr. 321, 209
\item[]
Schmutzer, E.: 2001, Astron. Nachr. 93, 103
\end{description}

\vspace{4ex}

\noindent Address of the author:
\\[2ex] Ernst Schmutzer
        Cospedaer Grund 57, D-07743 Jena  Germany

\end{document}